\documentclass[runningheads]{llncs}
\usepackage[T1]{fontenc}
\usepackage[table,xcdraw]{xcolor}
\usepackage{graphicx}
\usepackage{algorithm}
\usepackage{algorithmic}
\usepackage{amsmath} 
\usepackage{amsfonts} 
\usepackage{amssymb} 
\usepackage{bm} 
\usepackage{colortbl} 
\usepackage{booktabs}
\usepackage{multirow}
\usepackage{enumitem}
\definecolor{grayish}{RGB}{240, 240, 240} 
\definecolor{grayish_2}{RGB}{230, 230, 230} 
\usepackage{hyperref}
\usepackage{adjustbox}
\definecolor{tmi_blue}{cmyk}{100,0.37,0.0,0.15}
\hypersetup{
    colorlinks=true,
    linkcolor=tmi_blue,
    citecolor=tmi_blue,
    urlcolor=tmi_blue
}
\usepackage{color}
\usepackage{ulem}
\usepackage[misc]{ifsym}

\usepackage{subcaption}

\begin{document}
\title{RP-SAM2: Refining Point Prompts for Stable Surgical Instrument Segmentation}

\titlerunning{RP-SAM2}

\author{Nuren Zhaksylyk\inst{1}\textsuperscript{\href{mailto:nuren.zhaksylyk@mbzuai.ac.ae}{\Letter}}\and
Ibrahim Almakky\inst{1} \and
Jay Paranjape\inst{2} \and
S. Swaroop Vedula\inst{4} \and
Shameema Sikder\inst{3} \and
Vishal M. Patel\inst{2} \and
Mohammad Yaqub\inst{1}
}

\authorrunning{Nuren et al.}

\institute{%
    Mohamed bin Zayed University of Artificial Intelligence, Abu Dhabi, UAE \\
    \href{mailto:nuren.zhaksylyk@mbzuai.ac.ae}{\Letter} \space \email{nuren.zhaksylyk@mbzuai.ac.ae} \and
     Department of Electrical and Computer Engineering, The Johns Hopkins University, Baltimore, USA \and
    Wilmer Eye Institute, The Johns Hopkins University, Baltimore, USA \and
    Malone Center for Engineering in Healthcare, The Johns Hopkins University, Baltimore, USA
}


%
\maketitle              
\begin{abstract}
Accurate surgical instrument segmentation is essential in cataract surgery for tasks such as skill assessment and workflow optimization. However, limited annotated data makes it difficult to develop fully automatic models. Prompt-based methods like SAM2 offer flexibility yet remain highly sensitive to the point prompt placement, often leading to inconsistent segmentations. We address this issue by introducing RP-SAM2, which incorporates a novel shift block and a compound loss function to stabilize point prompts. Our approach reduces annotator reliance on precise point positioning while maintaining robust segmentation capabilities. Experiments on the Cataract1k dataset demonstrate that RP-SAM2 improves segmentation accuracy, with a 2\% mDSC gain, a 21.36\% reduction in mHD95, and decreased variance across random single-point prompt results compared to SAM2. Additionally, on the CaDIS dataset, pseudo masks generated by RP-SAM2 for fine-tuning SAM2’s mask decoder outperformed those generated by SAM2. These results highlight RP-SAM2 as a practical, stable and reliable solution for semi-automatic instrument segmentation in data-constrained medical settings. The code is available at \href{https://github.com/BioMedIA-MBZUAI/RP-SAM2}{https://github.com/BioMedIA-MBZUAI/RP-SAM2}.

\end{abstract}
\section{Introduction}
In cataract surgery, tracking and segmenting surgical instruments is crucial for evaluating surgeon precision, analyzing instrument‐tissue interactions, and optimizing workflows. Past advancements in deep learning have enabled accurate segmentation models in the medical domain \cite{medsam,nnunet,medsa,medsam2,transunet,swinunet,unet++,vnet}, with modern approaches, trained on a large corpus of data, capable of generalizing well to unseen classes without additional supervision \cite{sam,sam2,seem}. However, this task still poses a significant challenge in surgical contexts due to complex scenes with insufficient contrast, object occlusions, and lighting and motion-related artifacts.
Furthermore, acquiring high-quality, large-scale annotated datasets in this domain is both challenging and expensive due to the specialized expertise required, privacy constraints, and the time-consuming nature of manual labeling. As a result, the available annotated data is often limited in terms of both quantity and diversity. To address this gap, semi-automatic or interactive segmentation techniques have become a practical alternative, enabling efficient annotation with minimal user input \cite{scribbleprompt,simpleclick,mivos,focalclick,graco,adaptiveclick,cgam,bai2014error,chen2021conditional}. For instance, a few clicks or coarse scribbles, while leveraging robust pretrained models to refine segmentation automatically, can significantly speed up the labeling process.

Existing prompt-based segmentation models, such as the Segment Anything Model (SAM) \cite{sam} and its extended version trained on large-scale video datasets SAM2 \cite{sam2}, provide flexible ways to delineate objects through text prompts, masks, bounding boxes, or point prompts. However, while text-based prompts can be useful for certain natural imaging tasks, recent studies have shown that point prompts generally yield higher segmentation accuracy for SAM in medical imaging settings \cite{svdsam}. Meanwhile, bounding-box prompts often become ambiguous when multiple instruments are closely positioned or partially occlude one another, encompassing more than one object in a single box. Consequently, point prompts emerge as a more reliable and user-friendly choice for semi-automatic instrument segmentation.

\begin{figure}[t]
\centering
\includegraphics[width=\textwidth]{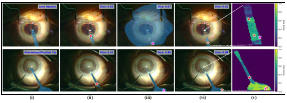}
\caption{Zero-shot performance of SAM2 on surgical instrument segmentation using single-point prompts. Rows represent instrument classes, with column (i) showing ground truth (GT) masks (blue shade). Columns (ii)-(vi) display single-point prompts (star symbol) and their segmentation masks, while column (v) presents a heatmap of dice scores for different point locations. Significant variance in results observed depending on prompt placement.} 

\label{fig:motiv_fig}
\end{figure}

Despite their convenience, point prompts contain an inherent flaw: even a slight change in the point location may dramatically alter the model’s output. Most existing models based on SAM2 focus on improving segmentation performance given high-quality prompts \cite{hqsam,medsam2,surgsam2}. As a result, these methods have limited consistency in the segmentation given point prompts from different regions of the object. A grid-based sampling of points across each instrument region in cataract surgery images shows significant fluctuations in the resulting prediction masks (Fig. \ref{fig:motiv_fig}). These findings highlight that SAM2’s segmentation performance heavily depends on the point location. In practice, suboptimal placements can yield poor masks, and adding more points does not always help; in certain cases, additional points may even degrade the final segmentation quality.

One solution would be to fine-tune SAM2 for surgical instrument segmentation. However, fine-tuning typically requires large amounts of annotated data, which are often unavailable in surgical settings, making this approach impractical. Some efforts \cite{robox-sam,pp-sam,stablesam} have focused on improving the stability of SAM given different numbers of points or noisy bounding boxes. More specifically, Stable-SAM \cite{stablesam} demonstrated instability in SAM's performance by averaging its performance scores across different numbers of points sampled from the object of interest and comparing them to its performance using a bounding box prompt. Stable-SAM then proposed a deformable sampling plugin that improves SAM's consistency by shifting the decoder's attention to the target region. Meanwhile, RoBox-SAM \cite{robox-sam} introduced a prompt refinement module to transform a low-quality box prompt into a high-quality one, and PP-SAM \cite{pp-sam} employed variably perturbed bounding box prompts during training to enhance SAM's robustness to noisy inputs. However, to the best of our knowledge, no publication has analyzed how shifting the same point across an object impacts the segmentation performance of SAM-based models. In this work, we address the underexplored problem of point prompt sensitivity in semi-automatic segmentation. Our contributions can be summarized as follows:
\begin{itemize}
    \item We introduce a novel shift block to SAM2 resulting in \textbf{R}obust \textbf{P}oint prompt SAM2 model, called \textbf{RP-SAM2}.
    \item We propose a compound loss for RP-SAM2 that enhances the segmentation performance regardless of the user's point prompt location.
    \item Our approach eases the annotator’s burden while maintaining the robust segmentation capabilities of foundation models such as SAM2.
\end{itemize}

\section{Methodology}
\textbf{Problem Formulation.}
A pre-trained foundation model for prompt based image segmentation can be defined as: \(S(I, P)=D\big(E(I),\xi(P)\big)\), where \(I \in \mathbb{R}^{H \times W \times 3}\) is the input image, \(P=(x,y)\) is the user's point prompt, \(E\) is the image encoder, \(\xi\) is the prompt encoder, and \(D\) is the mask decoder.  \(S(I,P)\) is trained to predict \(\hat{M}\), such that \(\hat{M} \approx M\), where \(M\) and \(\hat{M}\) are the true and predicted masks for an object of interest, respectively. As such, we also denote segmentation performance metric of \(S(I,P)\) as \(d=Dice(\hat{M},M)\). There exists a set of candidate points \(\mathbb{G} = \{G_i\}_{i=1}^N\) on the object, where each point \(G_i = (x^g_i, y^g_i)\) has corresponding prediction mask \(\hat{M_g}\) with \(d_{g}\geq d\). In this work, we propose RP-SAM2 with the goal of refining \(P\) into \(P^\prime=(x^\prime,y^\prime)\) with corresponding prediction mask \(\hat{M}^\prime\) to get \(d^\prime\geq d\). 

\begin{figure}[t]
\centering
\includegraphics[width=0.98\textwidth]{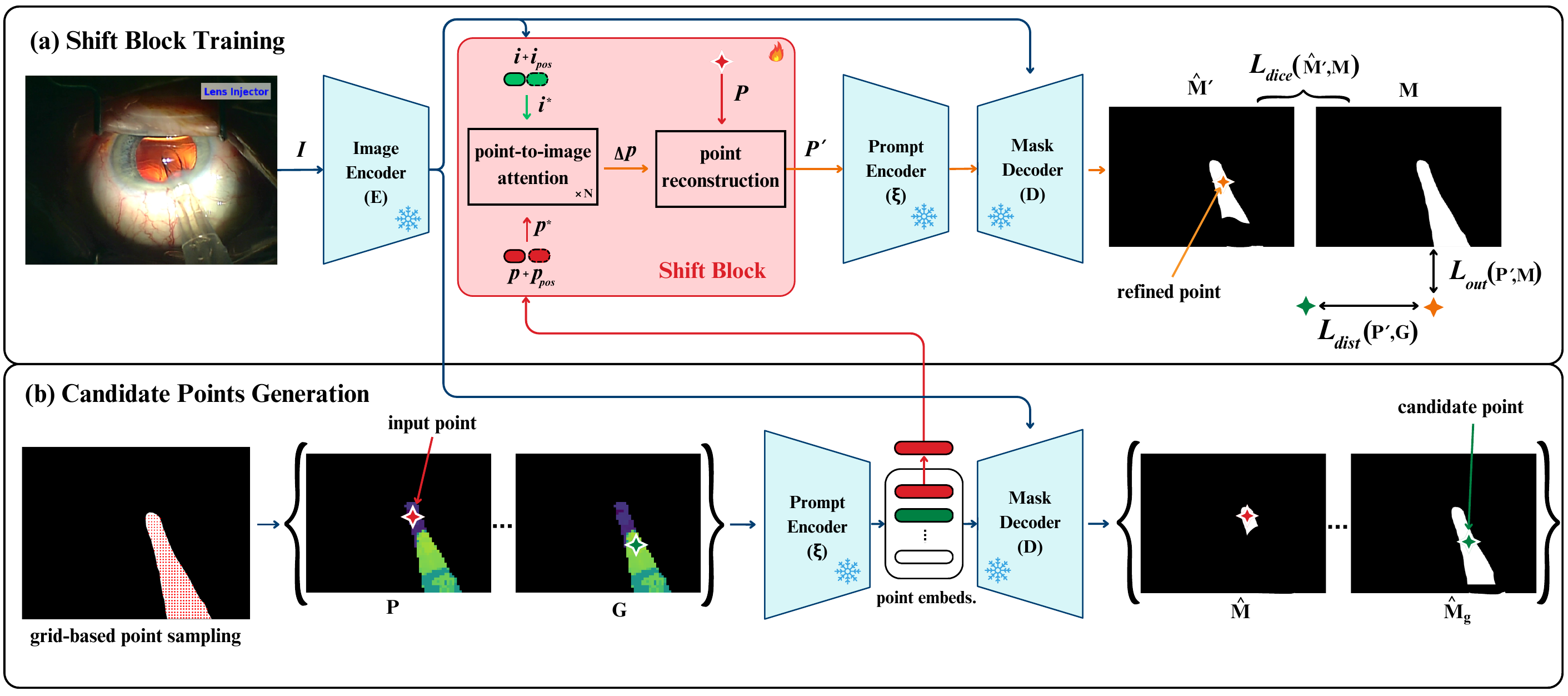}
\caption{Proposed RP-SAM2 architecture. (a) A Shift Block with 12.1M trainable parameters is integrated into SAM2 (with 236.5M frozen parameters) to reposition the user's input point prompt via cross-attention with image embeddings. (b) The algorithm employs grid-based point sampling to compute dice scores at multiple locations on the object and selects candidate point prompt coordinates.}
\label{fig:arch}
\end{figure}

\noindent \textbf{Candidate Points Generation.} As depicted in Fig. \ref{fig:arch}, we perform a grid-based point sampling on the object of interest and for each sampled point, we compute its segmentation performance \(d_i=Dice(\hat{M}_i,M)\). We then determine the maximum performance value \(d_{max}=\max_id_i\), and define the candidate set \(\mathbb{G}\) as the collection of points \(G_i\) satisfying \(d_{\text{max}} - \epsilon \leq d_i \leq d_{\text{max}}\) where \(\epsilon\) is a predefined tolerance. This selection ensures that \(\mathbb{G}\) includes only those points whose corresponding segmentation masks yield performance close to the best observed. This process happens on-the-fly during training.

\noindent \textbf{Shift Block.}
In its present state, SAM2 treats input point prompt independently from the image and the object of interest. Hence, it is susceptible to variations due to different points on the same object. This is amplified for surgical instruments, since they are mostly occluded by other instruments or organs. We hypothesize that adding dependence between the prompt and the image is an important step towards addressing this problem. Motivated by this, as depicted in Fig. \ref{fig:arch}, in RP-SAM2, we propose a novel shift block \(\mathcal{F}(P, i, p)\) that generates refined point prompts based on the image and original point prompts. Thus, the segmentation model is changed to: \(S(I,P)=D\big(E(I),\xi(\mathcal{F}(P, i, p))\big)\) where \(i=E(I)\) and \(p = \xi(P)\) are image and point embeddings, respectively. Similar to vision transformers \cite{ViT}, we add learnable positional embeddings \(p_{pos}\) and \(i_{pos}\) so the model can incorporate spatial information about the point and the image features. As a result, we get \(p^*=p+p_{pos} \in \mathbb{R}^{1 \times C}\) and \(i^* = i+i_{pos} \in \mathbb{R}^{(h \times w) \times C}\), where \(h\) and \(w\) are dimensions of the latent image. We then perform point-to-image cross attention to get point offset token \(\Delta p \in \mathbb{R}^{1 \times C}\):
\begin{equation}
 \Delta p = \frac{\text{exp}(Q(p^*) \cdot K(i^*)^T) }{\sum \text{exp}(Q(p^*) \cdot K(i^*)^T)}\cdot V(i^*),
\label{eq:cross_attention}
\end{equation}
where \(Q, K, V\) are query, key, and value embedding projection functions, respectively. Next, we decode \(\Delta p\) to reconstruct a refined point \(P^\prime\):
\begin{equation}
P^\prime = P +\sigma(s_x,s_y)\cdot\tanh(\delta(\Delta p)).
\label{eq:point_recon}
\end{equation}
Here, \(\sigma\) is the sigmoid function, \(\delta \) is a linear embedding function, and \((s_x, s_y)\) are learnable tokens that scale the shift in the \(x\) and \(y\) directions. We separate them because surgical instruments have non-uniform shapes, demanding independent offsets along each axis. 
\newline
\noindent \textbf{Compound Loss.} To optimize \(S(I,P)=D\big(E(I),\xi(\mathcal{F}(P, i, p))\big) \approx \hat{M}^\prime\), we freeze \(E,\xi, D\) and train only the shift block parameters using the following compound loss function:
\(
\mathcal{L}_{total} = \alpha \mathcal{L}_{dice} + \beta \mathcal{L}_{dist} + \gamma \mathcal{L}_{out},
\) where \(\mathcal{L}_{dist}(P^\prime, G_i)\) minimizes distance between the shifted point \(P^\prime\) and closest candidate \(G_i\), \(\mathcal{L}_{out}(P^\prime, M)\) restricts \(P^\prime\) to stay within object of interest, \(\mathcal{L}_{dice}(\hat{M}^\prime, M)\) maximizes the
overlap between predicted and true mask regions, and
\(\alpha, \beta, \gamma\) are hyperparameters set to control the weight of each loss component. 

\noindent \textbf{Dice Loss.}\ \(\mathcal{L}_{dice}(\hat{M}^\prime, M)\) measures how well the predicted refined segmentation mask \(\hat{M}^\prime\) aligns with the ground‐truth mask \(M\). Notably, it helps the shift block learn a more accurate prompt location by guiding the refined segmentation toward better overlap.

\noindent \textbf{Distance Loss.} To guide the shift block more effectively, we supervise it by providing multiple candidate points. To this end, we propose an exponential-based distance loss function, \(\mathcal{L}_{dist}\), to measure the discrepancy between each refined point and its closest candidate. Specifically, \(\mathcal{L}_{dist}\) imposes stronger penalties as the distance grows, thereby encouraging refined points to stay near their candidates. Formally, for each refined point \(P^\prime = (x^\prime, y^\prime)\), we compute its distance to every candidate point \(G_i = (x_i^g, y_i^g) \in \mathbb{G}\) as follows:
\begin{equation}
\mathcal{L}_{dist}(P^\prime, G_i)
= \frac{1}{2}
\Bigl[
\exp\bigl(\theta \cdot |x^\prime - x_i^g|\bigr)
+ \exp\bigl(\theta \cdot |y^\prime - y_i^g|\bigr)
- 2
\Bigr],
\label{eq:cross_attention}
\end{equation}
where \(\theta\) is a regularization term that exponentially penalizes larger deviations from a candidate. Finally, minimum distance across all \(G_i\) is taken as final loss. This approach avoids a purely deterministic strategy of focusing on a single target, since multiple candidates may yield equally high segmentation performance.

\noindent \textbf{Outside-Object Loss Function.} 
We propose an additional binary cross entropy-based loss,  \(\mathcal{L}_{out}\), to constrain \(P^{\prime}\) to lie within the target object’s boundaries. Since \((x^{\prime}, y^{\prime})\) are floating-point coordinates, we compute \(\hat{m}\) - bilinear interpolation of the mask values at neighboring pixels. If \(\hat{m} < 0.5\), the point is deemed outside the object, and we penalize it accordingly.

\section{Experimental Setup}
\textbf{Datasets.} 
We report results on the Cataract1k \cite{cataract} for in-distribution and CaDIS \cite{cadis} for out-of-distribution (OOD) analysis. In Cataract1k, there are 30 cataract surgery videos with 2256 frames and pixel-level annotations for 10 instrument classes. We use the shuffle 0, consisting of 2256 frames divided into training, validation, and test splits as provided in the dataset. We report results on the test split of CaDIS dataset (Task 2). Additionally, we fine-tuned the shift block on subsets of the training split, generated by stratified sampling to keep the class balance, for $50$ epochs each, and evaluated performance on the same test split.
\newline
\textbf{Implementation Details.} 
We use SAM2 with a Hiera-Large \cite{hiera} image encoder and integrate our shift block into it, keeping all SAM2 components frozen and only training our shift block. To replicate a real-world scenario, we do a grid sampling of points in each object and refine 10 random points during training. 
This approach reflects the variability of the user clicking different regions on the object of interest. We empirically set the hyperparameters, including the learning rate to 1e-4 and the batch size to 4. We kept all other hyperparameters and data augmentations the same as those used to train SAM2. We use 6 layers of point-to-image attention blocks. We initialize \((s_x, s_y)\) with 0.1 each, choose \(\epsilon\) as 0.5\%, and set \(\theta\) in \(L_{dist}\) to 15, \(\alpha, \beta, \gamma\) weights in \(L_{total}\) to 1.0, 0.1 and 1.0, respectively. We train shift block for 1500 epochs on NVIDIA RTX 5000 Ada and choose the best checkpoint based on validation results. 
\newline
\textbf{Evaluation Metrics.}
We compare all the models using the mean Dice Similarity Coefficient (mDSC) and mean 95th percentile Hausdorff Distance (mHD95). We report the average and standard deviation scores for 10 different points on the object generated by different random seeds. 
\section{Results and Discussion}
\begin{figure}[t]
    \centering
    \begin{subfigure}[b]{0.7\textwidth}
        \centering
        \includegraphics[width=\textwidth]{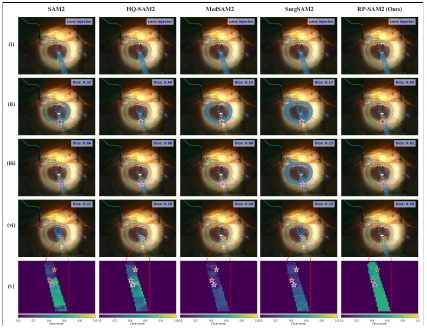}
        \caption{}
        \label{fig:qual_results}
    \end{subfigure}
    \hfill
    \begin{subfigure}[b]{0.281\textwidth}
        \centering
        \includegraphics[width=\textwidth]{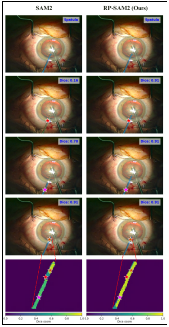}
        \caption{}
        \label{fig:noise_artefact}
    \end{subfigure}
    
    \caption{An illustration of (a) Qualitative comparison with SOTA and (b) An example of strong light-reflection artefacts on the instrument. Columns represent SOTA methods, row (i) GT masks for given instrument, rows (ii)-(vi) different single-point clicks on the instrument and corresponding predicted masks and (v) heatmap of dice scores for different single-point prompts.}
    \label{fig:combined_ood_ablation}
\end{figure}

\begin{table}[t]
\centering
\caption{Average performance and stability of SOTA methods (with corresponding trainable parameter counts) on the Cataract1k test set, evaluated over 10 single-point prompts. For each prompt, performance is first averaged across samples, and then the overall mean and standard deviation are computed across the prompts. Rows correspond to instrument classes, with bold indicating the best performance and underlined indicating the second best.}
\label{table:main_results}
\resizebox{\textwidth}{!}{
\begin{tabular}{c|ll|ll|ll|ll|ll} 
\toprule
\multirow{2}{*}{\textbf{Classes}} 
& \multicolumn{2}{c|}{\textbf{SAM2 \cite{sam2}} | 236.5M}
& \multicolumn{2}{c|}{\textbf{HQ-SAM2 \cite{hqsam}} | 5.1M}
& \multicolumn{2}{c|}{\textbf{MedSAM2 \cite{medsam2}} | 38.9M}
& \multicolumn{2}{c|}{\textbf{SurgSAM2 \cite{surgsam2}} | 11.5M}
& \multicolumn{2}{c}{\textbf{RP-SAM2 (Ours)} | 12.1M} \\
\cmidrule(lr){2-3} \cmidrule(lr){4-5} \cmidrule(lr){6-7} \cmidrule(lr){8-9} \cmidrule(lr){10-11} 
 & \multicolumn{1}{c}{\textbf{mDSC(\%)} $\uparrow$} & \multicolumn{1}{c|}{\textbf{mHD95} $\downarrow$} 
 & \multicolumn{1}{c}{\textbf{mDSC(\%)} $\uparrow$} & \multicolumn{1}{c|}{\textbf{mHD95} $\downarrow$} 
 & \multicolumn{1}{c}{\textbf{mDSC(\%)} $\uparrow$} & \multicolumn{1}{c|}{\textbf{mHD95} $\downarrow$} 
 & \multicolumn{1}{c}{\textbf{mDSC(\%)} $\uparrow$} & \multicolumn{1}{c|}{\textbf{mHD95} $\downarrow$} 
 & \multicolumn{1}{c}{\textbf{mDSC(\%)} $\uparrow$} & \multicolumn{1}{c}{\textbf{mHD95} $\downarrow$} \\
\midrule

\textbf{CC}
 & \underline{80.17}±\underline{3.05} & \underline{31.56}±\underline{9.94}
 & 66.37±4.13 & 64.73±18.58
 & 73.18±\textbf{2.67} & 49.76±12.59
 & 78.29±3.25 & 38.13±16.47
 & \textbf{81.16}±3.10 & \textbf{22.98}±\textbf{6.71}
 \\
\textbf{CF}
 & \underline{73.98}±\textbf{1.96} & \underline{47.28}±15.41
 & 51.97±3.06 & 106.27±27.73
 & 61.65±3.60 & 52.39±\textbf{13.90}
 & \textbf{76.76}±\underline{2.00} & \textbf{43.90}±\underline{15.21}
 & 73.40±3.63 & 51.14±16.07
 \\
\textbf{G}
 & \textbf{79.98}±1.79 & \underline{33.74}±6.92
 & 69.91±1.72 & 58.27±8.03
 & 72.89±2.27 & 43.9±7.86
 & \underline{78.75}±\underline{1.49} & \textbf{33.57}±\textbf{4.75}
 & 78.29±\textbf{1.18} & 41.45±\underline{5.97}
 \\
\textbf{IA}
 & \underline{78.00}±\underline{1.23} & \underline{66.93}±\underline{5.99}
 & 55.43±1.85 & 143.50±11.06
 & 62.81±1.45 & 88.90±6.01
 & 69.15±1.81 & 119.61±10.19
 & \textbf{81.42}±\textbf{0.88} & \textbf{55.39}±\textbf{4.31}
 \\
\textbf{IK}
 & \underline{89.14}±9.71 & 80.78±27.62
 & 75.20±\underline{7.63} & 59.01±\textbf{17.60}
 & 82.46±9.44 & \underline{39.52}±\underline{18.26}
 & 87.91±8.43 & \textbf{31.11}±24.70
 & \textbf{91.52}±\textbf{4.02} & 72.35±19.37
 \\
\textbf{KF}
 & 71.64±5.56 & 60.32±27.63
 & 70.46±4.13 & \underline{35.05}±\textbf{12.44}
 & \textbf{74.31}±\underline{3.18} & \textbf{26.93}±26.06
 & 70.16±4.39 & 145.90±32.54
 & \underline{73.12}±\textbf{2.45} & 59.17±\underline{12.46}
 \\
\textbf{LI}
 & \underline{67.52}±\underline{4.34} & \underline{137.26}±16.96
 & 52.60±6.18 & 162.47±\underline{12.72}
 & 48.65±7.83 & 160.68±28.09
 & 63.03±4.67 & 140.01±20.66
 & \textbf{75.41}±\textbf{3.41} & \textbf{109.19}±\textbf{11.11}
 \\
\textbf{PhT}
 & \underline{72.94}±2.25 & \underline{106.85}±8.01
 & 54.84±1.70 & 167.62±\underline{6.35}
 & 56.78±2.39 & 135.68±8.61
 & 62.45±\underline{1.53} & 179.61±7.79
 & \textbf{75.33}±\textbf{0.75} & \textbf{89.28}±\textbf{5.10}
 \\
\textbf{SK}
 & \underline{89.75}±3.03 & 67.88±75.68
 & 83.86±15.38 & 66.51±65.02
 & 86.88±4.92 & \underline{27.80}±47.58
 & 83.03±\underline{1.55} & 193.53±\underline{35.52}
 & \textbf{92.85}±\textbf{0.08} & \textbf{5.38}±\textbf{0.05}
 \\
\textbf{S}
 & 74.33±\underline{1.04} & 102.8±9.99
 & 73.45±\textbf{1.01} & \textbf{64.53}±6.22
 & 66.62±1.28 & 83.33±\textbf{4.04}
 & \textbf{76.57}±1.08 & \underline{69.53}±5.58
 & \underline{75.03}±1.38 & 71.96±\underline{5.48}
 \\

\midrule
\rowcolor{cyan!15} \textbf{Avg.}
 & \underline{77.74}±3.40 & \underline{73.54}±20.42
 & 65.41±4.68 & 92.80±18.57
 & 68.62±3.90 & 70.89±\underline{17.30}
 & 74.61±\underline{3.02} & 99.49±17.34
 & \textbf{79.75}±\textbf{2.09} & \textbf{57.83}±\textbf{8.66}
 \\

\bottomrule
\end{tabular}}
\end{table}
Table~\ref{table:main_results} summarizes the segmentation results on the Cataract1k test set for 10 single-point prompts sampled on the target instruments: Casulorhexis Cystotome (CC), Capsulorhexis Forceps (CF), Gauge (G), Irrigation-Aspiration (IA), Incision Knife (IK), Katena Forceps (KF), Lens Injector (LI), Phacoemulsification Tip (PhT), Slit Knife (SK) and Spatula (S). On average, our method achieves around 2\% gain in mDSC and 16-points (21.36\%) decrease in mHD95 compared to SAM2 across all instruments classes, outperforming all state-of-the-art (SOTA) methods. In particular, our approach demonstrates better stability (i.e., lower variance across prompts) compared to competing methods, as evident from the standard deviation values, ±2.09\% for mDSC and ±8.66 for mHD95. Among individual instrument classes, LI shows the largest gain of 7.89\% in mDSC compared to SAM2. We hypothesize that this is because the lens injector is mainly transparent and has fewer discernible regions for user clicks, making it more sensitive to point locations as shown in Fig. \ref{fig:qual_results}. We computed a p-value of 1.24e-05 for the differences in DSC scores between SAM2 and RP-SAM2, confirming that the observed improvements are statistically significant.
\begin{figure}[t]
    \centering
    \begin{subfigure}[b]{0.47\textwidth}
        \centering
        \includegraphics[width=\textwidth]{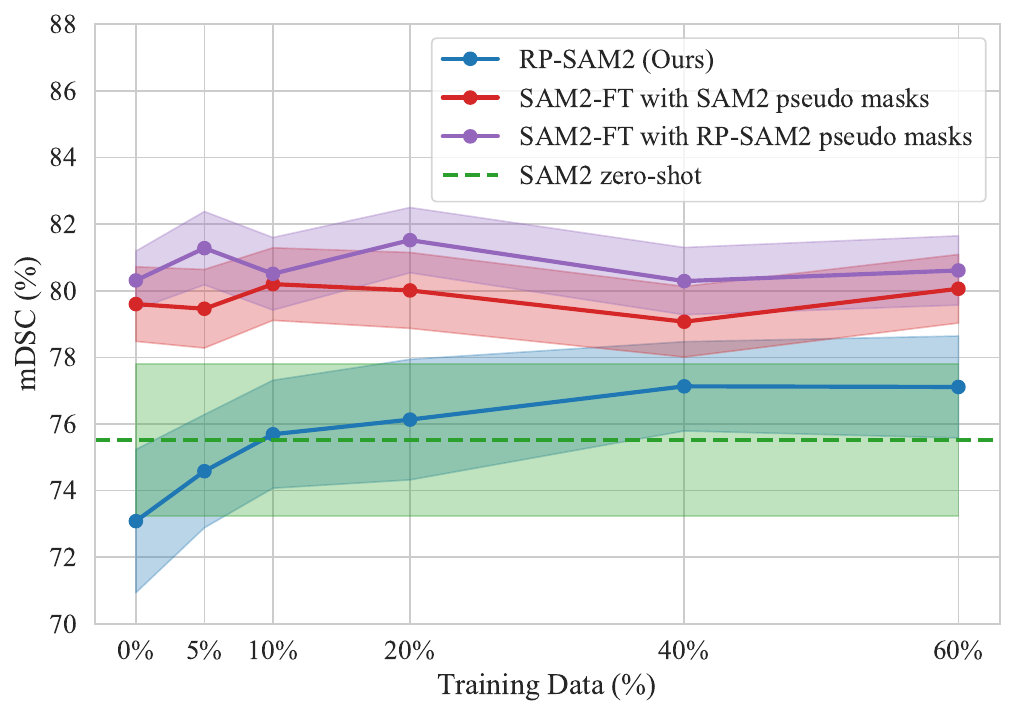}
        \caption{}
        \label{fig:ood_cadis}
    \end{subfigure}
    \hfill
    \begin{subfigure}[b]{0.47\textwidth}
        \centering
        \includegraphics[width=\textwidth]{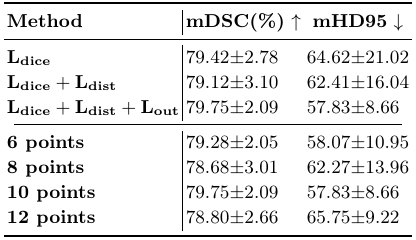}
        \caption{}
        \label{fig:ablation}
    \end{subfigure}
    
    \caption{An illustration of (a) OOD performance on the CaDIS test set, evaluated over 10 single-point prompts, where the vertical axis shows the mean dice score across instruments (with shaded standard deviation) and the horizontal axis indicates the percentage of the CaDIS training set used to fine-tune the shift block (with the remainder used to generate pseudo masks for fine-tuning SAM2’s mask decoder, referred to as SAM2-FT), (b) ablation study of loss components and the number of points sampled per object during training RP-SAM2 on Cataract1k dataset.}
    \label{fig:combined_ood_ablation}
\end{figure}

\noindent To assess generalizability, we conducted an OOD analysis on the CaDIS dataset (Fig. \ref{fig:ood_cadis}). Although our model did not surpass SAM2 in the zero-shot setting, using just 10\% of the CaDIS training data to fine-tune shift block allowed it to match SAM2’s mDSC, and employing 40\% resulted in significant improvements in both mDSC and stability. To demonstrate real-world applicability for semi-automatic labeling, we fine-tuned the mask decoder of SAM2 (hereafter SAM2-FT) using pseudo masks generated by both SAM2 and RP-SAM2. Specifically, we fine-tuned the shift block on 5-60\% of the training set and generated pseudo masks on the remaining 40-95\%, with the pseudo masks from RP-SAM2 yielding a higher performance boost. We find that although RP-SAM2 occasionally underperformed in the 0-5\% range - due to instances where points were shifted outside the instrument, leading to a zero dice score, and hence, a reduced mDSC - these cases were infrequent and did not significantly impact SAM2-FT’s overall performance. Still, this shows a possible area of improvement to motivate future research.
Regarding the ablation study, Fig. \ref{fig:ablation} shows that while combining \(L_{dist}\) with \(L_{dice}\) does not improve mDSC, it reduces mHD95, and adding \(L_{out}\) boosts both performance and stability on Cataract1k test set. Additionally, refining 10 points per instrument yields the best results, and even 6 or 8 points show improvement over SAM2.

\noindent \textbf{Qualitative Results.} Fig.~\ref{fig:qual_results} compares segmentation masks from various SOTA methods with RP-SAM2 model. Our model’s DSC heatmap (row v) is more uniform, ensuring consistent segmentation regardless of the selected point location, whereas SOTA methods have large DSC fluctuations with slight changes in point locations. Furthermore, our approach effectively handles light-reflection artifacts. For example, in Fig. \ref{fig:noise_artefact} (row ii), where SAM2 segments the reflection, our method shifts its target to recover the instrument boundary accurately. This behavior demonstrates greater robustness to noisy prompts and highlights that RP‐SAM2 leverages contextual cues to avoid mis-segmentation of reflective areas.

\section{Conclusion}
We present RP-SAM2, a refined version of SAM2 that is robust to user's point prompts for surgical instrument segmentation. RP-SAM2 shows improved stability, robustness to noise, and generalizability, coupled with quantifiable performance gains, which position RP-SAM2 as a valuable tool for advancing automatic surgical workflows and reducing annotation burdens in medical image analysis. The findings suggest that RP-SAM2 holds significant promise for practical applications in surgical image understanding. A potential future direction for improving RP-SAM2 could be to assess its performance in video segmentation by either tracking the position of adjusted prompts or recomputing optimal points at inference time.

\section{Acknowledgements}
JP, SSV, VP, and SK are supported by a grant from the National Eye Institute of the National Institutes of Health under award number R01EY033065. The content is solely the responsibility of the authors and does not necessarily represent the official views of the National Institutes of Health.

\bibliographystyle{splncs04}
\bibliography{arxiv}

\begin{thebibliography}{10}
\providecommand{\url}[1]{\texttt{#1}}
\providecommand{\urlprefix}{URL }
\providecommand{\doi}[1]{https://doi.org/#1}

\bibitem{ViT}
Alexey, D.: An image is worth 16x16 words: Transformers for image recognition at scale. arXiv preprint arXiv: 2010.11929  (2020)

\bibitem{bai2014error}
Bai, J., Wu, X.: Error-tolerant scribbles based interactive image segmentation. In: Proceedings of the IEEE Conference on Computer Vision and Pattern Recognition. pp. 392--399 (2014)

\bibitem{swinunet}
Cao, H., Wang, Y., Chen, J., Jiang, D., Zhang, X., Tian, Q., Wang, M.: Swin-unet: Unet-like pure transformer for medical image segmentation. In: European conference on computer vision. pp. 205--218. Springer (2022)

\bibitem{transunet}
Chen, J., Lu, Y., Yu, Q., Luo, X., Adeli, E., Wang, Y., Lu, L., Yuille, A.L., Zhou, Y.: Transunet: Transformers make strong encoders for medical image segmentation. arXiv preprint arXiv:2102.04306  (2021)

\bibitem{chen2021conditional}
Chen, X., Zhao, Z., Yu, F., Zhang, Y., Duan, M.: Conditional diffusion for interactive segmentation. In: Proceedings of the IEEE/CVF International Conference on Computer Vision. pp. 7345--7354 (2021)

\bibitem{focalclick}
Chen, X., Zhao, Z., Zhang, Y., Duan, M., Qi, D., Zhao, H.: Focalclick: Towards practical interactive image segmentation. In: Proceedings of the IEEE/CVF Conference on Computer Vision and Pattern Recognition. pp. 1300--1309 (2022)

\bibitem{mivos}
Cheng, H.K., Tai, Y.W., Tang, C.K.: Modular interactive video object segmentation: Interaction-to-mask, propagation and difference-aware fusion. In: Proceedings of the IEEE/CVF Conference on Computer Vision and Pattern Recognition. pp. 5559--5568 (2021)

\bibitem{stablesam}
Fan, Q., Tao, X., Ke, L., Ye, M., Zhang, Y., Wan, P., Wang, Z., Tai, Y.W., Tang, C.K.: Stable segment anything model. arXiv preprint arXiv:2311.15776  (2023)

\bibitem{cataract}
Ghamsarian, N., El-Shabrawi, Y., Nasirihaghighi, S., Putzgruber-Adamitsch, D., Zinkernagel, M., Wolf, S., Schoeffmann, K., Sznitman, R.: Cataract-1k dataset for deep-learning-assisted analysis of cataract surgery videos. Scientific data  \textbf{11}(1), ~373 (2024)

\bibitem{cadis}
Grammatikopoulou, M., Flouty, E., Kadkhodamohammadi, A., Quellec, G., Chow, A., Nehme, J., Luengo, I., Stoyanov, D.: Cadis: Cataract dataset for image segmentation. arXiv preprint arXiv:1906.11586  (2019)

\bibitem{robox-sam}
Huang, Y., Yang, X., Zhou, H., Cao, Y., Dou, H., Dong, F., Ni, D.: Robust box prompt based sam for medical image segmentation. In: International Workshop on Machine Learning in Medical Imaging. pp. 1--11. Springer (2024)

\bibitem{nnunet}
Isensee, F., Petersen, J., Klein, A., Zimmerer, D., Jaeger, P.F., Kohl, S., Wasserthal, J., Koehler, G., Norajitra, T., Wirkert, S., et~al.: nnu-net: Self-adapting framework for u-net-based medical image segmentation. arXiv preprint arXiv:1809.10486  (2018)

\bibitem{hqsam}
Ke, L., Ye, M., Danelljan, M., Tai, Y.W., Tang, C.K., Yu, F., et~al.: Segment anything in high quality. Advances in Neural Information Processing Systems  \textbf{36} (2024)

\bibitem{sam}
Kirillov, A., Mintun, E., Ravi, N., Mao, H., Rolland, C., Gustafson, L., Xiao, T., Whitehead, S., Berg, A.C., Lo, W.Y., et~al.: Segment anything. In: Proceedings of the IEEE/CVF International Conference on Computer Vision. pp. 4015--4026 (2023)

\bibitem{adaptiveclick}
Lin, J., Chen, J., Yang, K., Roitberg, A., Li, S., Li, Z., Li, S.: Adaptiveclick: Click-aware transformer with adaptive focal loss for interactive image segmentation. IEEE Transactions on Neural Networks and Learning Systems  (2024)

\bibitem{surgsam2}
Liu, H., Zhang, E., Wu, J., Hong, M., Jin, Y.: Surgical sam 2: Real-time segment anything in surgical video by efficient frame pruning. arXiv preprint arXiv:2408.07931  (2024)

\bibitem{simpleclick}
Liu, Q., Xu, Z., Bertasius, G., Niethammer, M.: Simpleclick: Interactive image segmentation with simple vision transformers. In: Proceedings of the IEEE/CVF International Conference on Computer Vision. pp. 22290--22300 (2023)

\bibitem{medsam}
Ma, J., He, Y., Li, F., Han, L., You, C., Wang, B.: Segment anything in medical images. Nature Communications  \textbf{15}(1), ~654 (2024)

\bibitem{vnet}
Milletari, F., Navab, N., Ahmadi, S.A.: V-net: Fully convolutional neural networks for volumetric medical image segmentation. In: 2016 fourth international conference on 3D vision (3DV). pp. 565--571. Ieee (2016)

\bibitem{cgam}
Min, S., Jeong, W.K.: Cgam: click-guided attention module for interactive pathology image segmentation via backpropagating refinement. In: 2023 IEEE 20th International Symposium on Biomedical Imaging (ISBI). pp.~1--5. IEEE (2023)

\bibitem{svdsam}
Paranjape, J.N., Sikder, S., Vedula, S.S., Patel, V.M.: S-sam: Svd-based fine-tuning of segment anything model for medical image segmentation. In: International Conference on Medical Image Computing and Computer-Assisted Intervention. pp. 720--730. Springer (2024)

\bibitem{pp-sam}
Rahman, M.M., Munir, M., Jha, D., Bagci, U., Marculescu, R.: Pp-sam: Perturbed prompts for robust adaption of segment anything model for polyp segmentation. In: Proceedings of the IEEE/CVF Conference on Computer Vision and Pattern Recognition. pp. 4989--4995 (2024)

\bibitem{sam2}
Ravi, N., Gabeur, V., Hu, Y.T., Hu, R., Ryali, C., Ma, T., Khedr, H., R{\"a}dle, R., Rolland, C., Gustafson, L., et~al.: Sam 2: Segment anything in images and videos. arXiv preprint arXiv:2408.00714  (2024)

\bibitem{hiera}
Ryali, C., Hu, Y.T., Bolya, D., Wei, C., Fan, H., Huang, P.Y., Aggarwal, V., Chowdhury, A., Poursaeed, O., Hoffman, J., et~al.: Hiera: A hierarchical vision transformer without the bells-and-whistles. In: International Conference on Machine Learning. pp. 29441--29454. PMLR (2023)

\bibitem{scribbleprompt}
Wong, H.E., Rakic, M., Guttag, J., Dalca, A.V.: Scribbleprompt: fast and flexible interactive segmentation for any biomedical image. In: European Conference on Computer Vision. pp. 207--229. Springer (2024)

\bibitem{medsa}
Wu, J., Ji, W., Liu, Y., Fu, H., Xu, M., Xu, Y., Jin, Y.: Medical sam adapter: Adapting segment anything model for medical image segmentation. arXiv preprint arXiv:2304.12620  (2023)

\bibitem{graco}
Zhao, Y., Li, K., Cheng, Z., Qiao, P., Zheng, X., Ji, R., Liu, C., Yuan, L., Chen, J.: Graco: Granularity-controllable interactive segmentation. In: Proceedings of the IEEE/CVF Conference on Computer Vision and Pattern Recognition. pp. 3501--3510 (2024)

\bibitem{unet++}
Zhou, Z., Rahman~Siddiquee, M.M., Tajbakhsh, N., Liang, J.: Unet++: A nested u-net architecture for medical image segmentation. In: Deep Learning in Medical Image Analysis and Multimodal Learning for Clinical Decision Support: 4th International Workshop, DLMIA 2018, and 8th International Workshop, ML-CDS 2018, Held in Conjunction with MICCAI 2018, Granada, Spain, September 20, 2018, Proceedings 4. pp. 3--11. Springer (2018)

\bibitem{medsam2}
Zhu, J., Qi, Y., Wu, J.: Medical sam 2: Segment medical images as video via segment anything model 2. arXiv preprint arXiv:2408.00874  (2024)

\bibitem{seem}
Zou, X., Yang, J., Zhang, H., Li, F., Li, L., Wang, J., Wang, L., Gao, J., Lee, Y.J.: Segment everything everywhere all at once. Advances in Neural Information Processing Systems  \textbf{36} (2024)

\end{thebibliography}

\end{document}